\documentclass{PoS}
\usepackage{amsmath}
\usepackage[utf8]{inputenc}
\def\beq{\begin{equation}}
\def\eeq{\end{equation}}
\def\Tr{{\rm Tr}}

\newcommand{\Ibb}{\ensuremath{\mathbb I} }
\newcommand{\cU}{\ensuremath{\mathcal U} }
\newcommand{\cUb}{\ensuremath{\overline{\mathcal U}} }

\title{The properties of $D1$-branes from lattice super Yang--Mills theory using gauge/gravity duality}
\ShortTitle{Lattice super Yang--Mills theory and gauge/gravity duality}
\author{Raghav Govind Jha \\
  Department of Physics, Syracuse University, Syracuse, NY 13244, USA \\
  E-mail: \email{rgjha@syr.edu} \\
}

\abstract{ 
The two-dimensional supersymmetric Yang-Mills (SYM) theory with sixteen supercharges at large $N$ and strong 't~Hooft coupling
is conjectured to be dual to certain supergravity solutions in the decoupling limit. We discretize the gauge theory
preserving a subset of supersymmetries on the lattice. Based on the choice of a point in the moduli space
for the expansion of the gauge links to target the correct continuum theory, one ends up with different lattice geometries. In our previous work, 
we explored the free energy and the phase structure on a skewed torus corresponding to $A_{2}^{*}$ lattice geometry. Here, we will consider 
square lattice and calculate the free energy, equation of state and speed of sound in this strongly coupled supersymmetric plasma. Since there is 
no shear viscosity in two dimensions, we comment on 
the expectations for the bulk viscosity from the calculations on the dual supergravity side, which unlike the conformal $\mathcal{N}=4$ SYM case 
does not vanish and is proportional to the trace of energy-momentum tensor. 
}

\FullConference{The 36th Annual International Symposium on Lattice Field Theory \\ Michigan State University, East Lansing, Michigan, USA \\  July 22-28, 2018}

\begin{document}
\setlength{\abovedisplayskip}{6 pt}
\setlength{\belowdisplayskip}{6 pt}

\section{Introduction}
It is conjectured that the strong coupling, large $N$ limit of supersymmetric theories possessing sixteen supersymmetries admit a holographic dual
\cite{Itzhaki:1998dd}. In the past couple of years, the program to access and understand the supergravity 
predictions using numerical simulations of supersymmetric gauge theories have evolved from its nascent 
stage and good agreement has
been observed. Several works \cite{Catterall:2008yz,Hanada:2008ez,Berkowitz:2016jlq} spread over the 
past decade have checked the thermodynamics
predicted from the supergravity side with the dual gauge theory observables with remarkable success in (0+1)-dimensions. 
The general form of the gauge/gravity duality is valid in lower dimensions as well, however, the 
four-dimensional case is special because $\mathcal{N}=4$ SYM theory is conformal. Lower dimensions are equally 
interesting because they can have a rich phase structure and are computationally cheaper to simulate using Monte Carlo methods.
Unlike SYM, the theory of strong interactions (QCD) has no known gravity dual, but QCD at high temperatures (about $T \ge 2.0-3.0~T_{c}$) is nearly a conformal field theory 
and is thought to be in the same universality class as $\mathcal{N}=4$ SYM. The thermodynamic potentials and transport coefficients have been calculated in 
QCD \cite{Boyd:1996bx, Nakamura:2004sy, Panero:2009tv} and relations to gravity dual via 
AdS/CFT have been explored.  Several important results in strongly coupled QCD have already been obtained using AdS/CFT conjecture, most famously 
the ratio of shear viscosity ($\eta$) to entropic density, $\eta/s$. In four dimensions, the only non-trivial 
viscosity coefficient is $\eta$ since the bulk viscosity ($\zeta$) vanishes in $\mathcal{N}=4$ SYM. 
Recently, we explored the two-dimensional supersymmetric gauge theory on a skewed torus and  
confirmed the phase transition between two different black hole solutions and computed the dual free energy in both phases \cite{Catterall:2017lub, Jha:2017zad}. 
In this proceedings, we \textit{propose} to study thermodynamics of the gauge theory in more detail by not only calculating the internal/free energy but 
rather the equation of state (EoS) on a square torus (hypercubic trajectories in the moduli space), which in turn will enable us to calculate the speed of the sound,
$\emph{i.e}$ $c_{s}$, the simplest transport coefficient.

\section{Theoretical background}
\subsection{Supergravity and its predictions}

\begin{figure}[htbp]
  \centering
      \includegraphics[height=2.51cm]{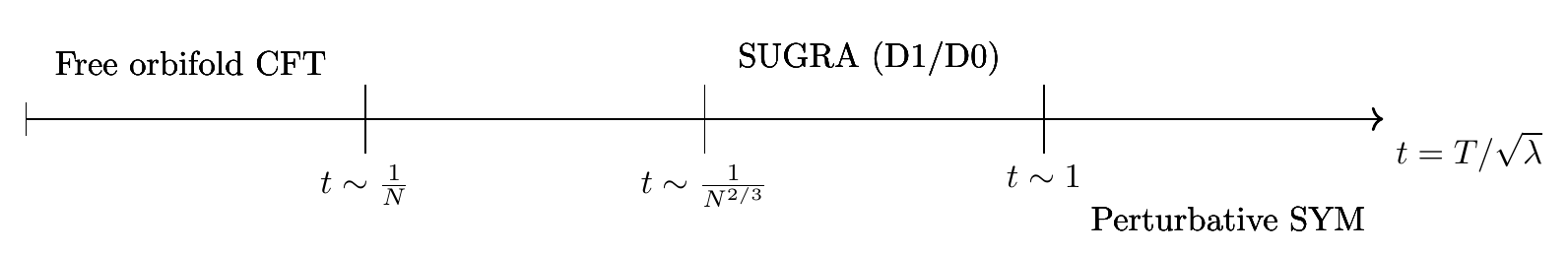}
  \caption{The different limits of the $\mathcal{N} = (8,8)$ SYM theory. We will focus on the region $1/N^{2/3} < t \ll 1$.}
  \label{fig:regime}
  \end{figure}

The IIB supergravity is dual to the `decoupling' limit of $N$ coincident $D1$-branes \cite{Itzhaki:1998dd}. In this limit, 
finite-energy excitations 
are considered simultaneously with the limits, 
$g_{\mathrm{YM}}^{2} = \frac{1}{2\pi} \frac{g_{\mathrm{s}}}{\alpha^{\prime}} = \mathrm{fixed}$ and $\alpha^{\prime}$ $\to$ 0, 
where $g_{\mathrm{s}}$ is the 
string coupling and $\alpha^{\prime}$ is the ~`Regge slope'. In the case of $D1$ branes, one starts out at weak coupling in the UV with 
a perturbative description. In the intermediate regime, there is supergravity (SUGRA) description in terms of $D0/D1$ brane solutions and at 
sufficiently low temperatures, one flows to a free orbifold CFT description. See Figure (\ref{fig:regime}) for a schematic representation of 
different regimes. 
The region in which the strongly coupled Yang Mills theory (denoting $p$ to be number of spatial dimensions) is 
dual to the Type IIA/IIB supergravity is given by, 
\beq\label{ineq:validiity} 
1 \ll \lambda_{\mathrm{eff}} \ll N^{\frac{10-2p}{7-p}} 
\eeq
where,  $\lambda_{\mathrm{eff}} = \lambda_{p+1} \beta^{3-p} = t^{-(3-p)}$, where $\lambda_{p+1}$ is the coupling in 
$(p+1)$-dimensions and $t$ is
the dimensionless temperature. This condition reduces to the familiar $1 \ll \lambda_{4} \ll N$ in four dimensions.
 We will refer to $\lambda_{2}$ and $\lambda$ interchangeably. 

Assuming the event horizon of the black hole geometry is at $ U = U_{0}$ (see \cite{Itzhaki:1998dd,Wiseman:2013cda} for details).
Then, we can calculate the temperature associated with the supergravity metric $T$ as,  
\begin{align}\label{eq:temp} 
T = \frac{(7-p) U_{0}^{\frac{5-p}{2}}}{4\pi \sqrt{d_{p} \lambda_{p+1}}}
\end{align}
The corresponding energy can be easily calculated and gives,  
\begin{equation}\label{eq:Energy}
\frac{E}{N^{2}}\Big|_{\mathrm{Dp-brane}} = \frac{(9-p)U_{0}^{7-p} L^{p}}{2^{11-2p} \pi^{\frac{13-3p}{2}} \Gamma\left(\frac{9-p}{2}\right) \lambda_{p+1}^{2}}  
\end{equation}
It can be further shown that, 
\begin{equation}
\frac{E}{S} = \left(\frac{9-p}{14-2p}\right) T ~~~~~ ;\ ~~~~~~ \\ 
E - P = \frac{4P}{5-p}
\end{equation}
where, $S$ is the entropy, $E$ is the energy, and $P$ is pressure times the volume ($pV$). The speed of sound $c_{s}$ is then, 
\begin{equation}
c_{s} = \sqrt{\frac{\partial P}{\partial E}} = \sqrt{\frac{5-p}{9-p}} 
\end{equation}
The hydrodynamical coefficients for general Dp-branes, with $p \ge 2$ was calculated in  ~\cite{Mas:2007ng}. 
However, the case $p=1$ is special. It is the only odd $p$, with $p  < 5$ which is not conformal. Also, there is no shear 
viscosity in two dimensions. For $D1$-branes, it was found in ~\cite{David:2009np} that the speed of sound is 
$c_{s} = \sqrt{\frac{1}{2}}$. The entropy density and bulk viscosity are given by \footnote{Note that there is a typo in 
Equation (5.7) of ~\cite{David:2009np}},  

\begin{equation}
s = \frac{2^4 \pi^{5/2} N^2 T^2}{3^3 \sqrt{\lambda}} ~~~~~ ;\ ~~~~~~ \\ 
\zeta = \frac{2^2 \pi^{3/2} N^2 T^2}{3^3 \sqrt{\lambda}} 
\end{equation}
and hence, the ratio $\zeta/s = 1/4\pi$ similar to four dimensions but with $\eta$ replaced by $\zeta$. 
One would ideally expect that $c_{s} = \sqrt{\frac{1}{2}}$ will be obtained for a conformal fluid in (2+1)-dimensions, 
so it is interesting that this result was obtained in a two-dimensional SYM theory in a regime where it is $\emph{not}$ conformal. 
This has been discussed in \cite{Kanitscheider:2009as} where it was argued that the hydrodynamical properties of non-conformal
branes is fully determined in terms of conformal hydrodynamics. The focus of the lattice calculations will be to calculate, $c_{s}$,
over the entire region, where the $D1$-description is valid and provide a numerical outlook on this issue. 

In the well-studied $p=0$ case, there is a single phase since the temporal direction corresponding to the 
black hole horizon is always \emph{deconfined}. For $p=1$, there is an intricate phase structure corresponding to topology changing 
transitions \cite{Gregory:1993vy} also known as the \emph{black hole/black string} transition. Using holography, 
this is conjectured to be dual to the \emph{deconfinement} transition in the gauge theory which for the two-dimensional SYM theory 
is expected to occur around  $r_{x}^2 = c_{\text{gravity}} r_{\tau}$ ($r_{x} = \sqrt{\lambda} L$, $r_{\tau} = \sqrt{\lambda} \beta$), 
and $c_{\text{gravity}} \approx 2.45$ for the square torus \cite{Catterall:2010fx, Dias:2017uyv}. 

Apart from the free/internal energy, EoS and $c_{s}$, there are other interesting observables to measure using 
lattice calculations to compare to
their corresponding gravity predictions. One of these include the Wilson loops proposed in 
~\cite{Rey:1998ik,Maldacena:1998im} for supersymmetric gauge 
theories which also include the contribution from the $(9-p)$ adjoint scalars ($\Phi$). It is defined as follows, 

\begin{equation}
W = \frac{1}{N} \mathrm{Tr} ~ \hat{P} \mathrm{exp} \Bigg[ \oint_C d\tau \Big (A_{\mu}(x) \dot{x}^{\mu} + \hat{\theta}^i|\dot{x}| \Phi_{i}(x)\Big)\Bigg] , 
\end{equation}
where \textbf{$\hat{\theta}$} is the unit vector and C is the contour which is parametrized by 
$x^{\mu}(\tau)$. It is normalized such that large $N$ limit is well-defined. 
We mention the prediction for this observable obtained using supergravity calculations for $ p < 3$, where 
only the $p=0$ case has yet been discussed using numerical simulations \cite{Hanada:2008gy}

\begin{itemize} 
\item $p = 0$ : $ \text{log} ~ \langle W \rangle = 1.89~t^{-3/5}$
\item $p = 1$ : $ \text{log} ~ \langle W \rangle = 1.54~t^{-1/2}$
\item $p = 2$ : $ \text{log} ~ \langle W \rangle = 1.15~t^{-1/3}$
\end{itemize}  
Generally, $\text{log} ~ \langle W \rangle \sim t^{-(3-p)/5-p}  \sim \lambda_{\mathrm{eff}}^{1/(5-p)}$. Note that
for $\mathcal{N} = 4$ SYM, this gives the $\sqrt{\lambda}$ dependence. 

\subsection{Finite temperature supersymmetric gauge theory} 

We consider the maximally supersymmetric Yang--Mills theory on two-torus ($S^{1}_{\beta} \times S^{1}_{L}$) with anti-periodic 
boundary conditions for the fermions along the time cycle ($\beta = 1/T$) and denote the trace of the energy-momentum
tensor (also known as `trace anomaly' or `interaction measure') by $ \Delta 
= E - P $, where $E$ and $P$ are defined as, 

\begin{equation}
E = T^2 ~ \frac{\partial \text{ln}~ Z}{\partial T}  \bigg\rvert_{V}  ~~~~~ ;\ ~~~~~~ \\ 
P = VT ~ \frac{\partial \text{ln}~ Z}{\partial V}  \bigg\rvert_{T}  \label{eq:3}
\end{equation}
Using the approximation for Eq.~(\ref{eq:3}) for homogeneous systems as, 

\begin{equation}
P \approx T ~ \text{ln}~ Z ,  
\label{eq:8}
\end{equation}
we can deduce an expression that relates the pressure to $\Delta$ given by, 
\begin{align}
\frac{\Delta}{T^2} &= \frac{E}{T^2} - \frac{P}{T^2} \nonumber  \\  
&= T \frac{\partial}{\partial T} \Big(\frac{P}{T^2}\Big) \label{eq:10}    
\end{align}
Integrating Eq.(\ref{eq:10}) gives, 
\begin{equation}
\frac{P(T)}{T^2} - \frac{P(T_{0})}{T_{0}^2} = \int_{T_{0}}^{T} dT^{\prime} \frac{1}{T^{\prime ~ 3}} \Delta(T^{\prime})
\label{eq:7}
\end{equation}
In principle, this relation will help us determine the EoS and the speed of sound. The range of temperatures ($t = T/\sqrt{\lambda}$ and we set $\lambda = 1$)
considered for the numerical integration would have to satisfy, $ 1/N < t < \alpha^2 / c_{\text{gravity}} $ with $ t \ll 1$.

\section{Lattice action}
The lattice $\mathcal{N} = 4$ super Yang-Mills action based on topological twisting formulation can be written as a sum of $\mathcal{Q}$-exact and 
$\mathcal{Q}$-closed terms. The details (and, devils) can be found elsewhere ~\cite{Catterall:2009it,Catterall:2014vka}. For a 
recent review, see ~\cite{Schaich:2018mmv}. The action is given by,  

\begin{align}
  S_{\rm exact} & = \frac{N}{4\lambda_{\mathrm{eff}}} \sum_{\textbf{n}} \mathrm{Tr}\bigg[-\mathcal{\overline{F}}_{ab}(\textbf{n})\mathcal{F}_{ab}(\textbf{n}) - \chi_{ab}(\textbf{n}) \mathcal{D}_{[a}^{(+)}\psi_{b]}^{\ }(\textbf{n}) - \eta(\textbf{n}) \mathcal{\overline{D}}_a^{(-)}\psi_a(\textbf{n}) \nonumber \\
  &\hspace{81mm} + \frac{1}{2}\left( \mathcal{\overline{D}}_a^{(-)}\mathcal{U}_a(\textbf{n})\right)^2\bigg],   \label{eq:exact} \\  
  S_{\rm closed} & = -\frac{N}{16\lambda_{\mathrm{eff}}} \sum_{\textbf{n}} \Tr \bigg[{\epsilon_{abcde}\ \chi_{de}(\textbf{n} + \hat{\mu}_a + \hat{\mu}_b + \hat{\mu}_c) \mathcal{\overline{D}}_c^{(-)} \chi_{ab}(\textbf{n})}\bigg],
\end{align}
where $\lambda_{\mathrm{eff}}$ is the dimensionless 't~Hooft coupling and we sum over repeated indices.
Supersymmetric theories have flat directions which are
a problem for numerical simulations and we control this by adding a $\mathcal{Q}$ breaking term to the lattice action (which we extrapolate to zero) as, 

\begin{equation}
  \label{eq:single_trace}
  S_{\text{flat}} = \frac{N}{4\lambda_{\mathrm{eff}}} \mu^2 \sum_{\textbf{n},\ a \neq 3} \Tr{\bigg(\cUb_a(\textbf{n}) \cU_a(\textbf{n}) - \Ibb_N\bigg)^2}
\end{equation}

Since we are interested in the dimensionally reduced $\mathcal{N} = 4$ SYM to two dimensions with the same number of supercharges 
(known as $\mathcal{N} = (8,8)$ SYM), we dimensionally 
reduce along the two spatial directions. To have a meaningful dimensional reduction, we have observed that an extra term has to be added to the action given by, 

\begin{equation}
  \label{eq:center}
  S_{\text{center}} = \frac{N}{4\lambda_{\mathrm{eff}}} \mu^2 \sum_{\textbf{n},\ i = x, y} \text{ReTr}\bigg[\bigg(\varphi_i(\textbf{n}) - \Ibb_N \bigg)^{\dagger}\bigg(\varphi_i(\textbf{n}) - \Ibb_N \bigg)\bigg].
\end{equation}

\begin{figure}
\begin{center} 
\includegraphics[width=0.72\textwidth]{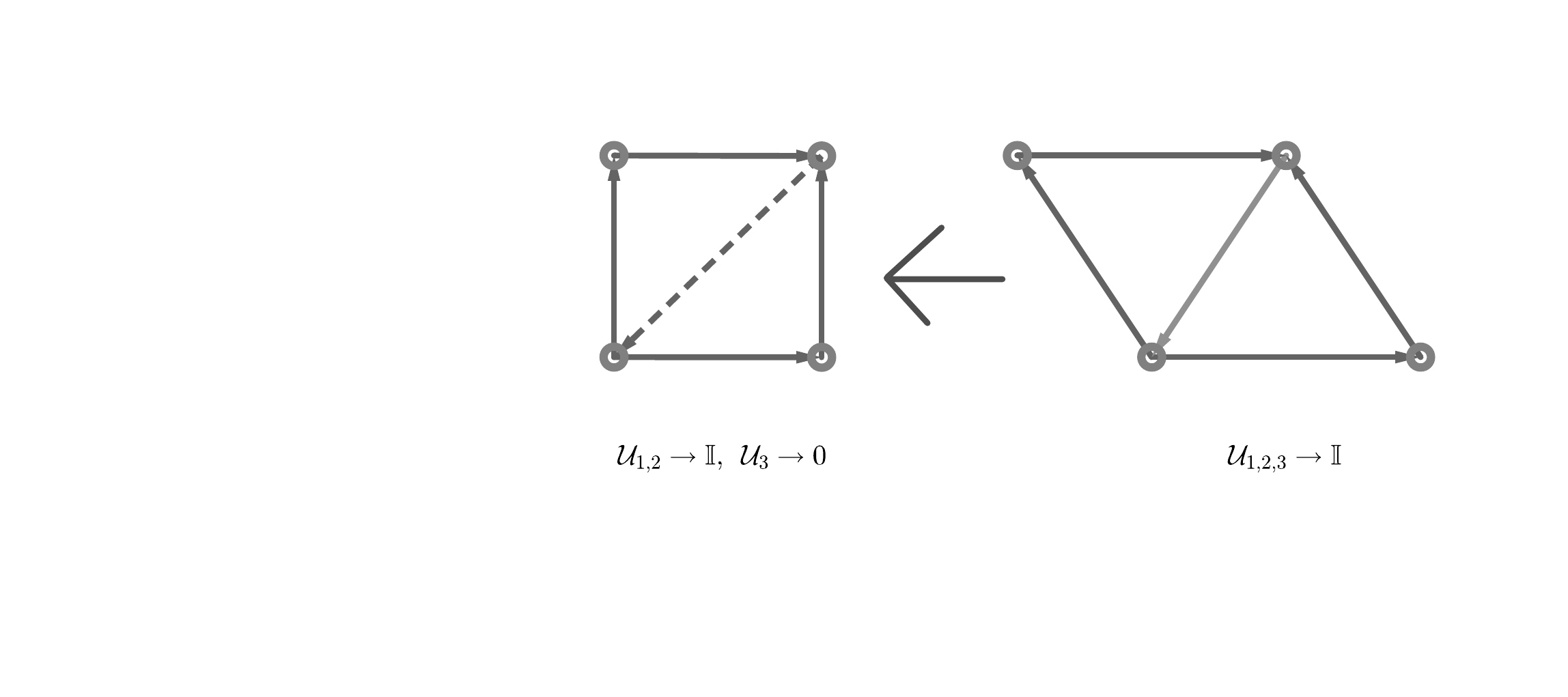}
\end{center}
\caption{\label{fig:plot}On the right we have $A_{2}^{*}$ lattice where the three links are treated equally and expanded symmetrically to 
target the continuum theory. On the left, by modifying the 
third link and requiring that it is expanded around zero, we get square lattice. See \cite{Kaplan:2005ta, Unsal:2005yh} for details.} 
\end{figure}

The lattice supersymmetric theories based on $\mathcal{Q}$-exact formulation are naturally adapted to non-orthogonal lattices. 
For example,  $\mathcal{N} = 4$ in four dimensions is formulated on $A_{4}^{*}$ lattice 
which has a bigger point group symmetry than the hypercubic lattice. However, we want to study the two-dimensional 
theory on a square lattice. In \cite{Kaplan:2005ta, Unsal:2005yh}, it was argued that one can get different lattice geometries by the choice of the expansion point for the
fields in the moduli space (the trajectory one follows to the infinity) . We add an additional term, $ S_{A_{2}^{*} \to \text{hyp.}}$ given by, 

\begin{equation}
  \label{eq:hypercubic}
  S_{A_{2}^{*} \to \text{hyp.}} = \frac{N}{4\lambda_{\mathrm{eff}}} \sigma^2 \sum_{\textbf{n}} \Tr{\bigg(\cUb_3(\textbf{n}) \cU_3(\textbf{n})\bigg)^2}
\end{equation}
to the action which consists of the gauge links in the extra direction of the skewed geometry. The resulting lattice 
is square [see Figure (\ref{fig:plot})] and we keep $\sigma = O(1)$ fixed for all couplings/temperatures.
The complete lattice action reads,  
\begin{equation}
S = S_{\mathrm{exact}} + S_{\mathrm{closed}} +  S_{\text{flat}} + S_{\text{center}}  + S_{A_{2}^{*} \to \text{hyp.}}
  \label{eq:full_act}
\end{equation}
We are carrying out the numerical simulations using \ref{eq:full_act} on the parallel software \href{https://github.com/daschaich/susy/}{{\texttt{SUSY LATTICE}}}
developed in \cite{Schaich:2014pda} and will report the results in the future. 

\vspace{12 pt}
\noindent {\sc Acknowledgments:}~It is a pleasure to thank Simon Catterall for the encouragement 
and guidance over the years. I am grateful to David Schaich and Toby Wiseman for 
numerous discussions. I thank Joel Giedt and Anosh Joseph for continuing collaboration 
on studies of supersymmetric theories on the lattice. This research was supported by the US Department of Energy (DOE), 
Office of Science, Office of High Energy Physics, under Award Number DE-SC0009998. 
The numerical computations are underway using DOE-funded USQCD facilities at Fermilab. 

\bibliographystyle{utphys}
\bibliography{lattice2018}
\end{document}